\documentclass[12pt]{article}
\usepackage[symbol*]{footmisc}
\usepackage{array}
\usepackage{amsmath}
\usepackage{amssymb}
\usepackage{amsfonts}

\def\be{\begin{equation}}
\def\ee{\end{equation}}
\def\ba{\begin{eqnarray}}
\def\ea{\end{eqnarray}}

\def\bq{\begin{quote}}
\def\eq{\end{quote}}

 at 10truept
\newcommand{\de}{\mathrm{d}}

\newcommand{\beq}{\begin{equation}}
\newcommand{\eeq}{\end{equation}}
\newcommand{\beqa}{\begin{eqnarray}}
\newcommand{\eeqa}{\end{eqnarray}}

\def\ltap{\ \raise.3ex\hbox{$<$\kern-.75em\lower1ex\hbox{$\sim$}}\ }
\def\gtap{\ \raise.3ex\hbox{$>$\kern-.75em\lower1ex\hbox{$\sim$}}\ }
\def\gl{\ \raise.5ex\hbox{$>$}\kern-.8em\lower.5ex\hbox{$<$}\ }
\def\roughly#1{\raise.3ex\hbox{$#1$\kern-.75em\lower1ex\hbox{$\sim$}}}

\begin{document}

\thispagestyle{empty}
\begin{flushright}
arXiv:0905.3391\\ May 2009
\end{flushright}
\vspace*{1cm}
\begin{center}
{\Large \bf Pauli-Fierz gravitons\\
\vspace{0.2cm}
on Friedmann-Robertson-Walker background}\\
\vspace*{1.5cm} {\large Luca Grisa\footnote{\tt
lgrisa@physics.umass.edu} and
Lorenzo Sorbo\footnote{\tt sorbo@physics.umass.edu}}\\
\vspace{.5cm}  
{\em Department of Physics, University of Massachusetts, Amherst, MA 01003}\\\vspace{.15cm} \vspace{1.5cm} ABSTRACT
\end{center}

We derive the Hamiltonian describing Pauli-Fierz massive gravitons on a flat Friedmann-Robertson-Walker (FRW) cosmology in a particular, non-generic effective field theory. The cosmological evolution is driven by a scalar field $\Phi$ with an arbitrary potential $V\left(\Phi\right)$. The model contains two coupled scalar modes, corresponding to the fluctuations of $\Phi$ and to the propagating scalar component of the Pauli-Fierz graviton. In order to preserve the full gauge invariance of the massless version of the theory, both modes have to be taken into account. We canonically normalize the Hamiltonian and generalize the Higuchi bound to FRW backgrounds. We discuss how this bound can set limits on the value of the Pauli-Fierz mass parameter. We also observe that on a generic FRW background the speed of propagation of the scalar mode of the graviton is always smaller than the speed of light.

\vskip2.5cm

\begin{flushleft}
%\hfill
%{PACS: }\\
%\keywords{}
%astro-ph/yymmnnn\\ September 2005
\end{flushleft}

\vfill \setcounter{page}{0} \setcounter{footnote}{0}
\newpage

%%%%%%%%%%%%%%%
\section{Introduction}%%%
%%%%%%%%%%%%%%%

The discovery of the recent cosmic acceleration in the late '90s has revamped the interest on theories that attempt to modify gravity at large distances.
An interesting area of studies is to allow for a small, but finite graviton mass. By assuming the mass to be sufficiently small, the gravitational interaction should depart from the predictions of the theory of General Relativity only at large distances, or equivalently late times, that is at scales comparable with the graviton Compton wave-length.

It was shown by van Dam and Veltman~\cite{vanDam:1970vg} and by Zakharov~\cite{Zakharov:1970cc} that this is in fact not the case on flat Minkowski background. For instance, for a mass of the order of the current value of the Hubble parameter, modifications are already manifest at Solar System size, where the contribution coming from the mass would have been expected to be irrelevant. The discontinuity, known as vDVZ~discontinuity, is only present for a flat background metric, for it was shown in~\cite{Porrati:2000cp,Kogan:2000uy} that the zero-mass limit is smooth when a non-zero cosmological constant is present. The removal of the vDVZ discontinuity in (Anti)-de Sitter space is due to the existence of two mass scales~-- the cosmological constant $H^2$ and the graviton mass $m^2$. However, for positive values of the cosmological constant, a new pathological regime appears: if the mass of the graviton does not obey the inequality $m^2\ge 2\,H^2$, known as the Higuchi bound~\cite{Higuchi:1986py,Deser:2001wx}, the theory develops a ghost-like instability.

In the light of the recent interest in potentially phenomenological implications of theories in which gravity is modified in the infrared regime, it is interesting to discuss how such a bound can be extended to more realistic backgrounds than pure de Sitter space. It is therefore the goal of the present work to investigate the effects of a Pauli-Fierz mass term for the graviton when the background space-time is described by a flat Friedmann-Robertson-Walker metric ({\em i.e.}, when the mass scale $H$ is time dependent) and, in particular, to generalize the Higuchi bound to a FRW Universe.\footnote{It is worth to mention that in a recent paper \cite{Gabadadze:2008ha} a mechanism was described to modify the Higuchi bound in a de Sitter Universe by adding a suitable coupled ghost; in our work we will consider only canonically normalized scalar fields. See also~\cite{Deffayet:2004ru} for the study of the behavior of a different model of infrared-modified gravity on FRW background.}

In order to consistently account for the dynamics of the energy density that drives the expansion of the Universe, we will consider a Pauli-Fierz graviton coupled to a scalar field $\Phi$ with a generic potential $V\left(\Phi\right)$. The background metric is characterized by a time-dependent Hubble scale $H(t)$. In order to maintain the full coordinate reparametrization invariance of the massless theory, we have to consistently take into account the fluctuations of $\Phi$ along with those of the graviton. This leads, in the scalar sector of the theory, to a complicated coupled dynamics. By following closely the analysis of~\cite{Deser:2001wx}, we find the generalized Higuchi bound to now depend -- through the behavior of $\Phi\left(t\right)$ -- not only on the Hubble scale $H(t)$ but also on its first time derivative $\dot{H}(t)$. This bound has to be satisfied at any time during the cosmological evolution for the theory not to develop any instabilities. If we assume the bound to be universal, that is to be independent on the species that drives the cosmological expansion, our result strongly constrains the value of the Pauli-Fierz mass.

Furthermore we will show that a physical degree of freedom of this model displays a Lorentz-violating propagation, that however never develops into superluminality.
\newline
\newline
The paper is organized as follows: in section 2 we will illustrate the Hamiltonian of the system describing gravity with a Pauli-Fierz mass term and a scalar field. In section 3 we will diagonalize the part of the Hamiltonian that describes the propagating tensor and vector modes. In section 4 we will discuss the conditions under which the scalar sector can be brought to canonical form (details can be found in the appendices) and, as usual, the most interesting dynamics occurs there. The implications of this non-trivial dynamics will be discussed in section~5.

%%%%%%%%%%%%%%%%%%%%%%%%%%%%%%%%%%%
\section{Quadratic Lagrangian of Massive Gravity on FRW}%%%
%%%%%%%%%%%%%%%%%%%%%%%%%%%%%%%%%%%%

Our system consists of a scalar field $\Phi$ canonically coupled to gravity, which we allow to slightly depart from standard GR.
\begin{equation}
\label{action}
S=S_0+S_m\,.
\end{equation}
$S_0$ is given by
\begin{equation}
S_0\equiv-\int d^4x\sqrt{-{}^{(4)}\!g}\ \left\{{}^{(4)}\!R-\frac{1}{2}\partial_\mu\Phi\,\partial^\mu\Phi-V\left(\Phi\right)\right\}\,,
\end{equation}
and $S_m$ is the most general quadratic self-interaction of the metric that does not exhibit (ghost-like) instabilities and preserves the Lorentz invariance of the Einstein-Hilbert action. The term is known in literature as the Pauli-Fierz mass term:
\begin{equation}
\label{PFaction}
S_m\equiv\int d^4x\,L_m=-\frac{m^2}{4}\,\int d^4x \sqrt{-^{(4)}\bar g}\, h_{\mu\nu}\,\left[\bar{g}^{\mu\rho}\,\bar{g}^{\nu\sigma}-\bar{g}^{\mu\nu}\bar{g}^{\rho\sigma} \right]\,h_{\rho\sigma}\,,
\end{equation}
where $\bar g_{\mu\nu}$ is the background metric that solves the equations of motion at zeroth order in the fluctuations $h_{\mu\nu}=g_{\mu\nu}-\bar g_{\mu\nu}$ over said background. By construction the Pauli-Fierz term does not contribute to the equations of motion for the background since it is quadratic in the fluctuations, therefore the metric $\bar g_{\mu\nu}$ is found by solving the standard Einstein equations of motion in presence of a scalar field $\Phi$.

Before proceeding any further it is worth to make few comments.
Firstly, in the present paper we will study the properties of the model~\eqref{action} in the spirit of effective field theory: we will be assuming that {\it a} UV completion does exist and reduces at low energies to a massive theory of gravity described by~\eqref{action} with mass given by~\eqref{PFaction}.

One might wonder why we constrain ourselves to a Lorentz-invariant mass term, while the background we are interested in is ultimately Lorentz-violating. As we will see in~section~\ref{Discussion}, the Lorentz-violating background will give rise to Lorentz-violating phenomena: in the high frequency regime, the speed of propagation of (the scalar component of) the gravitational modes is smaller than the speed of light. Hence it might seem natural to allow for a more general mass term, one that will only preserve the symmetries of a FRW background. The reason for not doing so is twofold: on one side, we would like to make contact with the literature on PF massive gravity -- the first and foremost motivation of this work was to extend the Higuchi bound to cosmological backgrounds. On the other end, we want to analyze what kind of Lorentz-violating effects arise because of a background that only preserves a subgroup of the symmetries of the theory; considering a more general quadratic self-interaction would only entangle those effects coming from a non-maximally symmetric background to those generated by a Lorentz-violating modification of the Einstein-Hilbert action.

The action can be rewritten in the ADM formalism as:
\begin{equation}
S_0=\int d^4x \left\{\pi^{ij}\,\dot g_{ij}+\Pi\,\dot\Phi+ \left[ N\  {\cal E}^0+N_i\ {\cal E}^i\ \right]\right\}\,,\\
\end{equation}
where ${\cal E}^0={\cal R}^0-{\cal T}^0$ and  ${\cal E}^i={\cal R}^i-{\cal T}^i$ and
\begin{eqnarray}
{\cal R}^0&=&\sqrt{g}\, R +\frac{\pi^{ij}\pi^{lm}}{\sqrt g}\,\left[\frac{1}{2}\,g_{ij}g_{lm}-g_{il}g_{jm} \right] \ ,\nonumber\\
{\cal T}^0&=&\frac{\Pi^2}{2\sqrt{g}}+\sqrt{g}\,\left[\frac{1}{2}g^{ij}\partial_i\Phi\partial_j\Phi+V\left(\Phi\right)\right]\,,\nonumber\\
{\cal R}^i&=&2\sqrt{g}\,D_j\left(\frac{\pi^{ij}}{\sqrt{g}}\right)\ , \qquad {\cal T}^i=\Pi\,\partial^i\Phi\,\,.
\end{eqnarray}
The following quantities, known as lapse and shift function, are usually not dynamical
\begin{equation}
N\equiv 1/\sqrt{-g^{00}}\, ,\quad N_i\equiv g_{0i}\,.
\end{equation}
The canonical momenta, $\pi^{ij}$ and $\Pi$, are related to the canonical coordinates, $g_{ij}$ and $\Phi$, via
\begin{eqnarray}
\dot{g}_{ij}&=&\frac{2\,N}{\sqrt g}\,\left(\pi_{ij}-\frac{1}{2}g_{ij}\,\pi\right)+D_{(i}\,N_{j)}\,,\nonumber\\
\dot\Phi&=&\frac{N\,\Pi}{\sqrt{g}}+N^i\,\partial_i\Phi\,.
\end{eqnarray}
The PF mass term can be rewritten in this formalism as
\begin{equation}
S_m=-\frac{m^2}{4}\,\int d^4x \,a^{-1}\,\left[ h_{ij}\,h_{ij}-h_{ii}\,h_{jj}-2\,a^2\,N_i\,N_i-4\,a^2\,n\,h_{ii} \right]\,.
\end{equation}

The background solution for $\Phi$ is denoted as $\bar\Phi(t)$ and depends only on time since we are interested in a cosmological background, which by construction does not depend on \eqref{PFaction}. This has the form of a flat Friedmann-Robertson-Walker, that is $\de\bar{s}_0^2=\bar{g}_{\mu\nu}\,\de x^\mu \de x^\nu=-\de t^2+a^2\left(t\right)\,\de\bf{x}^2$.

We then consider small perturbations over this background. We will study the dynamics of perturbations over this background. It is known that in theories of massive gravity there is an unexpected very low strong coupling scale, usually of order $\Lambda_5=(M_P\,m^4)^{1/5}$\ \footnote{We will conservatively estimate in the present work $\Lambda_5$ to be the strong coupling scale: the precise evaluation of this scale will be the subject of a future work.} and above such scale perturbative theory can not be trusted; the following analysis is therefore considered only within the range of validity of the effective theory, that is for wavelengths smaller than the Hubble radius and longer than the strong coupling scale. The study of the nonlinear effects on top of this theory is part of a future follow-up of this work.

Following the notation of~\cite{Deser:2001wx} the fields can be written as
\begin{eqnarray}
g_{ij}&=&\bar{g}_{ij}+h_{ij}=a^2\delta_{ij}+h_{ij}\nonumber\\
\pi^{ij}&=&\bar{\pi}^{ij}+p^{ij}=-2\,H\,a\,\delta^{ij}+p^{ij}\nonumber\\
N&=&1+n\,\,,\qquad \Phi=\bar{\Phi}+\varphi\,\,,\qquad\Pi=\bar{\Pi}+\pi\,\,.
\end{eqnarray}
$N_i$ is already a first order quantity, since its background value vanishes.

At quadratic level, the Lagrangian has the following form
\begin{equation}
L^Q=p^{ij}\,\dot{h}_{ij}+\pi\,\dot\phi+n\,{\cal E}^0_L+{\cal E}^0_Q+N_i\,{\cal E}^i_L+L_m
\end{equation}
where the subscripts $L$ and $Q$ denote linear and quadratic order of the quantities under consideration.

In the absence of the graviton mass term, both $n$ and $N_i$ appear linearly in the Lagrangian. Not being propagating degrees of freedom, their equations of motion enforce four constraints on the physical propagating modes. Hence the total number of physical degrees of freedom is two for the graviton, as expected from the Lorentz representation of a massless tensor field, and one for the scalar field.

When the mass term is introduced, the dynamics changes. The absence of time derivatives for $N_i$ still suggests that this mode is not propagating, but the mass term introduces a quadratic term for $N_i$: its equation of motion is therefore an algebraic constraint for $N_i$ itself. Two additional modes begin to propagate in this model, a transverse vector and a scalar field, accounting to five degrees of freedom for the graviton as demanded by the Lorentz representation of a massive tensor field.

The Lagrangian reduces to the following, once $N_i$ is integrated out
\begin{equation}\label{quadacfin}
L=p^{ij}\, \dot{h}_{ij}+\pi\,\dot\varphi +n\, ({\cal E}^0_L+\frac{m^2}{a}\,h_{ii})+{\cal E}_Q^0-\frac{\left({\cal E}^i_L\right)^2}{2\,a\,m^2}-\frac{m^2}{4\,a}\,
        \left(h_{ij}\,h_{ij}-h_{ii}\,h_{jj}\right)\,.
\end{equation}

Following \cite{Deser:2001wx}, the expression can be furthermore simplified through the redefinitions: $h_{ij}\rightarrow a^{1/2}\,h_{ij}$, $p^{ij}\rightarrow a^{-1/2}\,p^{ij}$, $n\rightarrow a^{-3/2} n$, $\varphi\rightarrow a^{-3/2}\,\varphi$, $\pi\rightarrow a^{3/2}\,\pi$. We also define $\Delta=\partial_i\partial_i/a^2$ (note that $\Delta$ depends on time) and rescale $\bar\Pi\rightarrow a^3\,\bar\Pi$ ({\it i.e.}, $\bar\Pi=\dot{\bar\Phi}$).

The background equations are
\begin{eqnarray}
&&H^2\equiv\,\frac{\dot{a}^2}{a^2}=\frac{\bar\Pi^2}{12}+\frac{\bar{V}}{6}\,\,,\nonumber\\
&&\dot{H}=-\frac{\bar\Pi^2}{4}\,\,,\nonumber\\
&&\dot{\bar\Pi}=-3\,H\,\bar\Pi-\bar{V}'
\end{eqnarray}
where $\bar{V}\equiv V(\bar{\Phi})$ etc.

Finally, we exploit the $SO(3)$ invariance of the background to decompose $h_{ij}$ in tensor, vector and scalar modes:
\begin{equation}
h_{ij}=h_{ij}^{Tt}+\partial_{(i} h_{j)}^t+\frac{1}{2}\,\left(\delta_{ij}-\frac{\partial_i\partial_j}{\partial_k\partial_k}\right)\,h^t+\frac{\partial_i\partial_j}{\partial_k\partial_k}\,h^l
\end{equation}
and analogously $p^{ij}$. Through this definition we can now study the scalar, vector and tensor components of the action separately.

%%%%%%%%%%%%%%%%%%%%%
\section{Tensor and Vector Modes}%%
%%%%%%%%%%%%%%%%%%%%%

The Lagrangian for the tensor modes $h_{ij}^{Tt}$ takes the form
\begin{equation}
L_{\mathrm {tensor}}=p^{Tt}_{ij}\,\dot{h}^{Tt}_{ij}-\left[\left(p^{Tt}_{ij}-\frac{5}{4}\,H\,h^{Tt}_{ij}\right)^2+\frac{1}{4}\,h^{Tt}_{ij}\,\left(-\Delta+m^2-\frac{9}{4}\,H^2-4\,\dot{H}\right)\,h^{Tt}_{ij}\right]\,\,,
\end{equation}
that is easily brought to canonical form by defining $p^{Tt}_{ij}\rightarrow p_{ij}/\sqrt{2}+5\,H\,q_{ij}/(2\,\sqrt{2})$ and $h^{Tt}_{ij}\rightarrow \sqrt{2}\,q_{ij}$. The final form of the Lagrangian is thus
\begin{equation}
L_{\mathrm {tensor}}=p_{ij}\,\dot{q}_{ij}-\left[\frac{1}{2}\,p_{ij}\,p_{ij}+\frac{1}{2}\,q_{ij}\,\left(-\Delta+m^2-\frac{9}{4}\,H^2-\frac{3}{2}\,\dot{H}\right)\,q_{ij}\right]\,\,,
\end{equation}
and we see that tensor modes have always well-behaved kinetic terms and, as usual, obey the same equation of motion of minimally coupled scalars.

The Lagrangian of the vector modes, $h_i^t$, can be written as 
\begin{eqnarray}
L_{\mathrm {vector}}=2\,p_i\,\dot{q}_i&-&\left[2\,p_i\,\left(1-\frac{\Delta}{m^2}\right)\,p_i+\frac{1}{2}q_i\,\left(m^2+4\,H^2-16\,H^2\,\frac{\Delta}{m^2}+\bar\Pi^2\right)\,q_i\right.+\nonumber\\
&+&\left. H\,p_i\left(-5+8\,\frac{\Delta}{m^2}\right)\,q_i\right]
\end{eqnarray}
where we have defined $p_i=\sqrt{-\partial_i\partial_i}\,p^t_i$ and $q_i=\sqrt{-\partial_i\partial_i}\,h^t_i$. As in~\cite{Deser:2001wx}, we perform the following canonical transformation
\begin{equation}
p_i\rightarrow \frac{4\,H\,p_i-\left(m^2-6\,H^2\right)\,q_i}{2\,m}\,\,,\qquad q_i\rightarrow \frac{3\,H\,q_i+2\,p_i}{2\,m}\,\,,
\end{equation}
that brings the Hamiltonian in canonical form
\begin{equation}
L_{\mathrm {vector}}=p_{i}\,\dot{q}_{i}-\left[\frac{1}{2}\,p_{i}\,p_{i}+\frac{1}{2}\,q_{i}\,\left(-\Delta+m^2-\frac{9}{4}\,H^2+ \frac{3}{2}\,\dot{H}\right)\,q_{i}\right]\,\,.
\end{equation}
Like for the tensor modes, also the vectors always present a well-behaved Hamiltonian (as long as $m\neq 0$ -- of course, for $m=0$ the vector modes turn into purely gauge modes).

%%%%%%%%%%%%%%%
\section{Scalar Modes}%%
%%%%%%%%%%%%%%%

Our starting Lagrangian has the form
\begin{equation}
L_{\mathrm scalar}=p^l\,\dot{h}^l+\frac{p^t\,\dot{h}^t}{2}+\pi\,\dot\varphi-{\cal H}_0\left(p^t,\,p^l,\,\pi,\,h^t,\,h^l,\,\varphi,\,n\right)\,\,.
\end{equation}

Since $n$ appears in the Lagrangian as a Lagrange multiplier, its contribution disappears once the corresponding constraint equation ${\cal E}^0_L+m^2\,a^{-1/2}\,h_{ii}=0$  is integrated. For scalar modes, such an equation reduces to
\begin{equation}
\nu^2\,\left(h^l+h^t\right)-\Delta\,h^t-2\,H\,\left(p^l+p^t\right)- \bar{V}'\,\varphi -\bar\Pi\,\pi=0\,\,,
\end{equation}
where we have defined the quantity 
\begin{equation}\label{constlapse}
\nu^2\equiv m^2-2\,H^2+\bar\Pi^2/2=m^2-2\,H^2-2\,\dot{H}=m^2-2\,\frac{\ddot{a}}{a}\,\,,
\end{equation}
that reduces to the Deser and Waldron parameter $\nu^2$ in the limit $\dot{H}\rightarrow 0$~\cite{Deser:2001wx}.

We use eq.~(\ref{constlapse}) to eliminate $p^t$ from our Lagrangian. Through several integrations by parts the Lagrangian can be written as
\begin{equation}\label{l2}
L_{\mathrm scalar}=p_0\,\dot{q}_0+p_1\,\dot{q}_1-{\cal H}_1\left(p_0,\,p_1,\,q_0,\,q_1,\,h^t\right)
\end{equation}
where we have defined the new canonical variables
\begin{eqnarray}
p_0\equiv p^l-\frac{\nu^2}{4\,H}\,h^t\,\,,&&\qquad p_1\equiv \pi+\frac{\bar{V}'}{4\,H}\,h^t\,\,,\nonumber\\
q_0\equiv h^l-\frac{h^t}{2}\,\,,&&\qquad q_1\equiv \varphi-\frac{\bar\Pi}{4\,H}\,h^t\,\,.
\end{eqnarray}

The variable $q_0$ is analogous to the one defined in~\cite{Deser:2001wx}.
The variable $q_1$ corresponds (modulo an overall factor $a\left(t\right)$) to the Mukhanov variable $v$~\cite{Mukhanov:1990me}, that is the canonically normalized scalar degree of freedom in a Universe filled by a scalar field. These are the two fundamental scalar degrees of freedom of our system.

The variable $h^t$ is not a dynamical degree of freedom, as it lacks its own canonical momentum $p^t$. Variation of the action with respect to $h^t$ gives an algebraic equation for $h^t$ whose solution reads
\begin{equation}\label{ht}
h^t=\frac{2}{3\, m^2\, \nu^2}\,\left[-m^2 \nu^2\,q_0+\left( \bar\Pi\,p_1 +\bar V'\,q_1\right) m^2-4\, H\, \Delta\,(\,p_0-H\,q_0-\frac{\bar\Pi}{2}\,q_1)\right]\,\,.
\end{equation}

Note that the above solution is singular for $\nu^2=0$. As discussed in~\cite{Deser:1983mm,Deser:2001wx,Deser:2001us}, the critical line $\nu^2=0$ corresponds to a partially massless theory, where $h^t$, appearing only linearly in the Lagrangian, is a Lagrange multiplier. As a consequence, the corresponding equation for $\nu^2=0$ provides one additional constraint that removes the scalar mode of the graviton from the propagating degrees of freedom. The result is that~-- at quadratic level~-- the graviton is left with only four helicities ($\pm 2,\,\pm 1$). As we will discuss in section 5, the possibility of living on the critical line $\nu^2=0$ is much less interesting in the generic FRW background considered here, since the line $\nu^2=0$ can be crossed, during the evolution of the Universe,  only when system is already in a unstable region of the parameter space.
 
Let us now go ahead and consider a generic $\nu^2\neq 0$. By plugging the result~(\ref{ht}) back into~(\ref{l2}) we obtain the final Hamiltonian. Since its form is rather complicated, here we will express it in the following compact notation
\begin{equation}\label{l0}
L_{\mathrm scalar}=p^T\cdot \dot{q}-\left(\frac{1}{2}\,p^T\cdot K\cdot p+\frac{1}{2}\,q^T\cdot M\cdot q+\,p^T\cdot V\cdot q\right)
\end{equation}
where we have defined the vectors $p^T\equiv\left(p_0,\,p_1\right)$ and $q^T\equiv\left(q_0,\,q_1\right)$ and where the explicit expression of the matrices $K$, $M$ and $V$ are given in Appendix A.

By performing a series of canonical transformations (details in Appendix A) we can bring the Hamiltonian of our system to the form
\begin{eqnarray}\label{semifinal}
{\cal {H}}_{\mathrm s}&=&\frac{m^2\,\lambda^4}{24\,H^2\nu^2}P_0^2-2\frac{H^2\left(\nu^2+2\,\lambda^2\right)}{m^2\,\lambda^4}Q_0\,\Delta Q_0+\frac{P_1^2}{2}-\frac{Q_1\,\Delta Q_1}{2}+\nonumber\\
&&-2 H\,\bar\Pi\frac{\nu^2}{\lambda^4}\,Q_0P_1+\frac{Q_i\,{M}_{ij}\,Q_j}{2}
\end{eqnarray}
where we have defined a new function of the background quantities
\begin{equation}
\lambda^2\equiv m^2-2\,H^2\,\,,
\end{equation}
and where the elements of ${M}_{ij}$ can be obtained by setting $\Delta=0$ in eqs.~(\ref{m00sm}-\ref{m11sm}) in Appendix A and are complicated functions of the background quantities.

Even if eq.~(\ref{semifinal}) does not yet correspond to the Hamiltonian in its canonical form $\vec{P}^2/2-\vec{Q}\Delta\vec{Q}/2+V(\vec{Q})$, it already shows an important peculiarity of our system. We see that the sign  of the coefficient of the kinetic term of the gravitational mode $P_0^2$ is that of $\nu^2$. This observation allows us to obtain a first extension of the Higuchi condition $m^2>2\,H^2$ to a generic cosmological background. In a cosmological background, we notice that a necessary condition for the positiveness of the Hamiltonian reads $\nu^2>0$, {\it i.e.},
\begin{equation}\label{ineq}
m^2>2\,H^2+2\,\dot{H}\,\,.
\end{equation}
If the above inequality were to be violated, then the kinetic term of the degree of freedom $Q_0$ would have the wrong sign, signaling an instability of the theory.

From now on we will assume that the inequality~(\ref{ineq}) is satisfied. Then, we can canonically normalize the variable $Q_0$ through the redefinition $P_0\rightarrow P_0\,\gamma-\dot\gamma\,Q_0$, $Q_0\rightarrow Q_0/\gamma$ where $\gamma\equiv 2\,H\nu\sqrt{3}/(m \lambda^2)$. We thus get the final result
\begin{eqnarray}\label{final}
{\cal {H}}_{\mathrm s}&=&\frac{P_0^2}{2}-\frac{1}{2}\,\left(\frac{1}{3}+\frac{2}{3}\,\frac{\lambda^2}{\nu^2}\right)\,Q_0\,\Delta Q_0+\frac{P_1^2}{2}-\frac{Q_1\,\Delta Q_1}{2}+\nonumber\\
&&-\bar\Pi\frac{m\,\nu}{\lambda^2\sqrt{3}}\,Q_0\,P_1+\frac{Q_i\,{M}_{ij}\,Q_j}{2}
\end{eqnarray}
where $M_{00}$ has a new form, given in eq.~(\ref{m00new}) in Appendix A,~$M_{01}$ is given by the same $M_{01}$ of eq.~\eqref{m01sm} divided by $\gamma$, and $M_{11}$ is unchanged.

Eq.~(\ref{final}) is our final expression for the Hamiltonian governing this system. In principle, one can perform a canonical transformation to eliminate the $Q_0P_1$. This is however not very illuminating, so we will relegate to Appendix B the discussion of the strategy that allows to eliminate such a term. 

Before discussing our results, we check that in the limit of de Sitter background ({\it i.e.}, $\bar\Pi$, $\bar{V}'\rightarrow 0$) we recover the result of~\cite{Deser:2001wx}. Indeed, in this limit the Hamiltonian reduces to
\begin{equation}
{\cal {H}}_{\mathrm s}^{dS}=\frac{P_0^2}{2}+\frac{P_1^2}{2}+\frac{Q_0}{2}\,\left(-\Delta+m^2-\frac{9}{4}\,H^2\right)\,Q_0+\frac{Q_1}{2}\,\left(-\Delta+\bar{V}''-\frac{9}{4}\,H^2\right)\,Q_1
\end{equation}
Of course, this canonical form can be obtained only if $\nu^2>0$, that is when the Higuchi bound $m^2>2\,H^2$ is satisfied for a Pauli-Fierz graviton on a de Sitter background.

%%%%%%%%%%%%%%%%%%%%%%%%%%%%%
\section{Discussion}%%%%%%%%
%%%%%%%%%%%%%%%%%%%%%%%%%%%%%
\label{Discussion}

The Hamiltonian for a massive graviton on a cosmological background governed by a scalar field has two remarkable properties. First, in order for the kinetic term of the gravitational modes to have the right sign, the system must obey the inequality $\nu^2>0$. Second, since the background breaks the isometry group down to $SO(3)$, the scalar mode of the graviton turns out to inherit a Lorentz non-invariant dispersion relation.

We will assume the condition $\nu^2>0$ to be met in the following discussion.

For wavelengths shorter than the characteristic timescale of the background and shorter than the Compton wavelength of the graviton, the mode $Q_0$ follows the equation
\begin{equation}\label{eqq}
\ddot{Q}_0-\left(\frac{1}{3}+\frac{2}{3}\frac{\lambda^2}{\nu^2}\right)\,\Delta\,Q_0=0\,\,.
\end{equation}
The coefficient of the Laplacian in the previous equation has to be positive to prevent the generation of exponentially increasing solutions. The explicit expression of this coefficient is
\begin{equation}\label{propspeed}
\frac{1}{3}+\frac{2\,\lambda^2}{3\,\nu^2}=1+\frac{4\,\dot{H}}{3\,\nu^2}=\frac{m^2-2\,H^2-2\,\dot{H}/3}{\nu^2}\,\,.
\end{equation}

Before discussing the stability of the model, let us note that the speed of propagation of these modes is $\sqrt{1+4\dot{H}/3\nu^2}$. Since $\dot{H}<0$, it is guaranteed that these modes do not experience superluminal propagation. Interestingly, \cite{Moore:2001bv} has observed that the speed of propagation of gravitational modes cannot be smaller than (at least) $(1-2\times 10^{-15})\,c$. Were this bound violated, gravitational Cherenkov radiation would have depleted the population of high energy cosmic rays that we currently observe. Unfortunately, this bound cannot be directly applied to our Pauli-Fierz gravitons, in fact it relies on the tensorial coupling of the graviton, while we were interested in the couplings of the scalar component of the Pauli-Fierz graviton. Moreover, the Pauli-Fierz theory is known to be strongly coupled at a very low scale of the order of $m\,\left(M_P/m\right)^{1/5}$~\cite{ArkaniHamed:2002sp}, while the bound mentioned above relies on the emission of gravitons with energies of the order of $10^{11}$~GeV. It would be interesting to study whether a bound, analogous to that of~\cite{Moore:2001bv}, could be applied to the scalar component of the Pauli-Fierz graviton.

The requirement of positivity of eq.~\eqref{propspeed}, {\em i.e.}, $m^2>2\,H^2+2\,\dot{H}/3$, provides a more restrictive condition than $\nu^2>0$. By considering the equation of state parameter of the background $w$ with $-1\le w\le 1$, $\dot H$ and $H^2$ are related as $\dot{H}=-3\,(1+w)\,H^2/2$.  The condition $\nu^2>0$ can be written as $m^2/H^2>-(1+3\,w)$, whereas the condition $(1/3+2\,\lambda^2/3\,\nu^2)>0$ is equal to $m^2/H^2>1-w$. It is easy to see that the second condition is always stronger than the first one and that they coincide only on a de Sitter background. As a consequence, the lower limit on the mass of the graviton on a cosmological background is given by
\begin{equation}\label{bound}
m^2>\left(1-w\right)\,H^2\,\,.
\end{equation}

We note that, since the requirement of positivity of eq.~\eqref{propspeed} is more restrictive than the bound $\nu^2>0$, the ``partially massless" regime discussed in~\cite{Deser:1983mm,Deser:2001us} can not be attained as a consequence of the cosmological evolution. While for a de Sitter background $\nu^2$ is a constant, in our case it is time-dependent, so $\nu^2$ might cross zero during the cosmological evolution. However, when this happens, the quantity~\eqref{propspeed} is already negative, so the theory is already in an unstable regime hence such a limit can not be trusted.

Assuming that the constraint~\eqref{bound} is independent of the kind of matter that is driving the cosmological expansion, we can infer a phenomenological bound on the mass of the Pauli-Fierz graviton. Indeed, we know that the Universe had a standard cosmological evolution at least from the time of nucleosynthesis, when it was in a thermal bath at a temperature of $\sim 10$~MeV. If we assume the mass of the graviton to be constant during the cosmological history of the Universe, then eq.~\eqref{bound} constrains the Pauli-Fierz mass to be larger than the value of the Hubble parameter at nucleosynthesis, $m\gtrsim 3\times 10^{-14}$~eV. Such a mass corresponds to a Compton wavelength of the order of tens of thousand of kilometers, which is ruled out since the solar system dynamics is well described by newtonian gravity. Of course, the limit $m\gtrsim 3\times 10^{-14}$~eV does not imply that a massless graviton is ruled out. Indeed, as we have discussed in section 2, we expect our linearized theory to break down for distances smaller than the scale at which the theory becomes strongly coupled. For values of $m$ so small that the strong coupling scale is comparable to the Hubble radius (and {\em a fortiori} for a massless graviton), we will not be able to trust our linearized analysis.

To conclude, we have derived the canonical Hamiltonian describing Pauli-Fierz massive gravitons on a FRW background driven by a scalar field. The scalar sector contains a mode with Lorentz non-invariant dispersion relation. The requirement that the momentum part of the Hamiltonian is positive definite induces the bound~\eqref{ineq}, whereas the (more restrictive) bound~\eqref{bound} prevents the development of instabilities through the generation of large gradients of the fields. Our formulation provides a setting for the study of the presence or the absence of the vDVZ discontinuity, as well as the existence of strongly coupled regimes, on a FRW background. We hope to go back to these points in a forthcoming publication.

\vspace{1cm}

{\bf \noindent Note added}

\smallskip

While we were at the final stages of the writing of this paper, the work~\cite{Blas:2009my}, whose subject partially overlaps with ours, was posted on the archive. The generalized Higuchi bound obtained with the Pauli-Fierz choice of parameters in~\cite{Blas:2009my} does not agree with our eqs.~\eqref{ineq} or \eqref {bound}, due to different nature of the starting Lagrangians.

\vspace{1cm}

{\bf \noindent Acknowledgements}

\smallskip

We thank John Donoghue for suggesting the problem and, with Nemanja Kaloper, for useful discussions. This work is partially supported by the U.S. National Science Foundation grant PHY-0555304.

\appendix

%%%%%%%%%%%%%%%%%%%%%%%%%%%%%
%%%%%%%%%%%%%%%%%%%%%%%%%%%%%
\section{Canonical Transformations}%%%
%%%%%%%%%%%%%%%%%%%%%%%%%%%%%
%%%%%%%%%%%%%%%%%%%%%%%%%%%%%

In this appendix, we show explicitly the canonical transformations that allow to bring the Lagrangian~(\ref{l0}) to the Hamiltonian~(\ref{semifinal}). Our procedure follows rather closely the one of~\cite{Deser:2001wx}. Our starting point is the Lagrangian~(\ref{l0}) 
\begin{equation}
L_{\mathrm {scalar}}=p^T\cdot \dot{q}-\left(\frac{1}{2}\,p^T\cdot K\cdot p+\frac{1}{2}\,q^T\cdot M\cdot q+\,p^T\cdot V\cdot q\right)\,\,,
\end{equation}
where
\begin{eqnarray}
&&K=\left(
\begin{array}{cc}
 3-\frac{4\, \Delta}{m^2}+\frac{4\, \Delta^2}{3\, m^2\,\nu^2} & \frac{\bar\Pi}{6 H}\, \left(3-\frac{2 \Delta }{\nu^2}\right) \\
  \frac{\bar\Pi}{6 H}\, \left(3-\frac{2 \Delta }{\nu^2}\right) &1+\frac{m^2\,\bar\Pi^2}{12\,H^2\,\nu^2}
\end{array}
\right)
\end{eqnarray}
\begin{eqnarray}
&&V=\left(
\begin{array}{cc}
 -\frac{\nu^2}{2\,H}-\frac{5}{2}\,H-\frac{\Delta}{3\,H}+4\,H\,\frac{\Delta}{m^2}-\frac{4}{3}\,\,\frac{\Delta^2}{m^2\,\nu^2} & 2\,\frac{\Delta\,\bar\Pi}{m^2}+\frac{\bar{V}'}{2\,H}-\frac{\Delta\,\bar{V}'}{3\,H\,\nu^2}-\frac{2}{3}\,\frac{\Delta^2\,\bar\Pi}{m^2\,\nu^2} \\ \bar\Pi\,\left(-1-\frac{m^2}{12\,H^2}+\frac{\Delta}{3\,\nu^2}\right)
 & \frac{3}{2}\,H+\frac{m^2\,\bar{V}'\,\bar\Pi}{12\,H^2\,\nu^2}+\frac{\bar\Pi^2\,\Delta}{6\,H\,\nu^2}
\end{array}
\right)
\end{eqnarray}
\begin{eqnarray}\scriptsize{
M=\left(
\begin{array}{cc}
 m^2+\Pi
   ^2-\frac{4\,\Delta\,H^2}{m^2}+\frac{4\,\Delta^2\,H^2}{3\,m^2\,\nu ^2}-\frac{2\,\Delta }{3}+\frac{m^2\,\nu ^2}{12\,H^2} & \left(\frac{2\,\Delta }{3\,\nu^2}-\frac{m^2}{6\,H^2}\right)\bar{V}'+\left(\frac{4\,H\,\Delta^2}{3\,m^2\,\nu ^2}-\frac{\Delta }{3\,H}-\frac{4\,H\,\Delta }{m^2}\right)
  \Pi  \\
\left(\frac{2\,\Delta }{3\,\nu^2}-\frac{m^2}{6\,H^2}\right)\bar{V}'+\left(\frac{4\,H\,\Delta^2}{3\,m^2\,\nu ^2}-\frac{\Delta }{3\,H}-\frac{4\,H\,\Delta }{m^2}\right)
  \Pi     & -\Delta+\bar{V}''-\frac{\Delta\,\bar\Pi^2}{m^2}\, 
   \left(1-\frac{\Delta }{3\,\nu ^2}\right) +\frac{\bar{V}'}{3 H \nu ^2}\,
   \left(\frac{\bar{V}'\,m^2}{4\,H}+\Delta\,\Pi \right)
\end{array}
\right)}\nonumber\\
\end{eqnarray}

First, we rewrite it in the form
\begin{equation}
L_{\mathrm {scalar}}=p^T\cdot \dot{q}-\left(\frac{1}{2}\,(p-h\cdot q)^T\cdot K\cdot (p-h\cdot q)+\frac{1}{2}\,q^T\cdot \tilde{M}\cdot q+(p-h\cdot q)^T\cdot \tilde{V}\cdot q\right)
\end{equation}
where the matrix 
\begin{eqnarray}\label{h}
h=\left(
\begin{array}{cc}
 H			& \bar\Pi/2 \\
 \bar\Pi/2	& -\bar{V}'/\bar\Pi-3\,H\,\left(1+\frac{\bar\Pi^2}{2\,\lambda^2}\right)
\end{array}
\right)
\end{eqnarray}
is chosen in such a way that the matrices $\tilde{M}\equiv M+h^T\cdot K\cdot h+h^T\cdot V+V^T\cdot h$ and $\tilde{V}\equiv V+K\cdot h$ acquire a simpler form. Note that we have defined the new time-dependent function
\begin{equation}
\lambda^2\equiv m^2-2\,H^2\,\,.
\end{equation}

The explicit expression of the matrices $\tilde{M}$ and $\tilde{V}$ is such that
\begin{eqnarray}
\begin{array}{c}
\tilde{M}+\dot{h}=\left(
\begin{array}{cc}
 \frac{m^2\,\lambda^4}{12\,\nu^2} & 0 \\ 0
 & -\Delta-\frac{3\,\bar\Pi}{4\,\lambda^4}\,\left[8\,H^2\,\lambda^2\,\bar{V}'+\bar\Pi\,\left(m^4+6\,H^2\,\nu^2-2\,H^2\,\bar\Pi^2\right)\right]
\end{array}
\right)\,\,,\nonumber\\
\,\,\nonumber\\
\tilde{V}=\left(
\begin{array}{cc}
 -\frac{m^2}{2\,H}+\frac{3}{2}\,H+\frac{\Delta}{3}\,\frac{\lambda^2}{\nu^2} & \bar\Pi\,\frac{4\,\Delta-3\,\bar\Pi^2}{4\,\lambda^2} \\ -\frac{m^2\,\lambda^2\,\bar\Pi}{12\,H^2\,\nu^2}
 & -\frac{\bar{V}'}{\bar{\Pi}}-H\,\left(\frac{3}{2}+\frac{2\,\bar\Pi^2}{\lambda^2}\right)
\end{array}
\right)\,\,\,.
\end{array}
\end{eqnarray}
We perform the following canonical transformation
\begin{eqnarray}
\begin{array}{c}
p=P+h\cdot Q+h\cdot\alpha\cdot P\,\,,\qquad q=Q+\alpha\cdot P\,\,,\\
\,\,\\
\alpha\equiv\left(\begin{array}{cc}
 \frac{H}{2}\,\frac{\left[4\,\lambda^4\,\left(3\,m^2-2\,\Delta-9\,H^2\right)-\left(12\,\bar{V}'+54\,H\,\bar\Pi\right)\,H\,\bar\Pi\,\lambda^2-3\,\left(m^2+2\,H^2\right)\,\bar\Pi^4\right]}{m^2\,\lambda^6} & \bar\Pi/\lambda^2 \\ \bar\Pi/\lambda^2  & 0
\end{array}
\right)\,\,,
\end{array}
\nonumber
\end{eqnarray}
followed by the simple canonical transformation $P_0\rightarrow Q_0$, $Q_0\rightarrow -P_0$. The Lagrangian now has the same form as in eq.~(\ref{l0}) where now
\begin{eqnarray}
\begin{array}{c}
K=\left(\begin{array}{cc}
\frac{m^2\,\lambda^4}{12\,H^2\,\nu^2} & 0 \\ 0 & 1
\end{array}
\right)\,\,,\\
V=\left(\begin{array}{cc}
\frac{\left(m^2+2\,H^2\right)\,\bar\Pi^2}{4\,H\,\lambda^2}+\frac{2\,H\,\bar\Pi^2+\bar{V}'\,\bar\Pi}{2\,\nu^2} & 0 \\ 
-2\,H\,\bar\Pi\,\nu^2/\lambda^4 & -\bar{V}'/\bar\Pi-H\,\left(3/2+2\,\bar\Pi^2/\lambda^2\right)
\end{array}
\right)\,\,,\\
M=\left(\begin{array}{cc}
M_{00} & M_{01}\\ 
M_{01} & M_{11}
\end{array}\right)\,\,,
\end{array}
\end{eqnarray}
with
\begin{eqnarray}
M_{00}&=&-\frac{4\,H^2\,\left(\nu^2+2\,\lambda^2\right)}{\lambda^4}\,\Delta+\frac{3\,H^2}{8\,m^2\,\lambda^8\,\nu^2}\,\left[16\,\lambda^4\,\Pi^2\,\nu^2\,\bar{V}''-32\,\lambda^6\,\bar{V}'{}^2+\right.\\
&&-\left. 16\,H\,\lambda^2\,\bar\Pi\,\left(12\,\lambda^2\,\nu^2+\bar\Pi^4\right)\,\bar{V}'+8\,\left(4\,m^2-9\,H^2\right)\lambda^8+12\,\left(m^2-40\,H^2\right)\lambda^6\,\bar\Pi^2\right]\,\,,\nonumber\\
M_{11}&=&-\Delta+\frac{3\,\bar\Pi}{4\,\lambda^4}\,\left[8\,H\,\lambda^2\,\left(\bar{V}'+2\,H\,\bar\Pi\right)+m^4\,\bar\Pi+6\,H^2\,\bar\Pi^3\right]\,\,,\\
M_{01}&=&-\frac{3\,H\,\bar\Pi^2}{\lambda^6}\,\left[2\,\left(\bar{V}'+H\,\bar\Pi\right)\,\lambda^2+3\,H\,\bar\Pi\,\nu^2\right]\,\,,
\end{eqnarray}

We can eliminate the diagonal terms of the matrix $V$ by $P\rightarrow P+A\cdot Q$,~~ $Q\rightarrow Q$ where the matrix $A$ is defined as ${\mathrm {diag}}(-V_{00}/K_{00},-V_{11}/K_{11})$. This leaves the matrix $K$ unchanged, while $V$ simplifies to
\begin{eqnarray}
V=\left(
\begin{array}{cc}
0 & 0 \\
-2\,H\,\bar\Pi\,\nu^2/\lambda^4 & 0
\end{array}\right)\,\,,
\end{eqnarray}
and the matrix $M$ has components
\begin{eqnarray}
M_{00}&=&-4\,\frac{H^2\,\left(\nu^2+2\,\lambda^2\right)}{m^2\,\lambda^4}\,\Delta+6\,\frac{H\,\bar\Pi}{m^2\,\lambda^6}\,\left[2\,m^2\,\nu^2-\lambda^2\,\left(m^2+3\,H^2\right)\right]\,\bar{V}'+\nonumber\\
&&+\frac{3\,H^2}{4\,m^2\,\lambda^8}\,\left[4\,\left(4\,m^2-9\,H^2\right)\,\lambda^6+2\,\left(11\,m^2-39\,H^2\right)\,\lambda^4\,\bar\Pi^2+\right.\nonumber\\
&&+\left.\left(19\,m^4 - 44\,m^2\,H^2 +\,12\,H^4\right)\,\bar\Pi^4 + 6\,m^2\,\bar{\Pi}^6\right]\,\,,\label{m00sm}\\
M_{01}&=&M_{10}=-2\,\frac{H}{\lambda^4}\,\left(5\,\nu^2-4\,\lambda^2\right)\,\bar{V}'-\frac{H^2\,\bar\Pi}{2\,\lambda^6}\,\left(6\,\lambda^4+35\,\lambda^2\,\bar\Pi^2+5\,\bar\Pi^4\right)\,\,,\label{m01sm}\\
M_{11}&=&-\Delta+\bar {V}''-\frac{9}{4}\,H^2-\frac{2\,H\,\bar{V}'\,\bar\Pi}{\lambda^2}+\frac{3}{8}\,\frac{\left(m^2-10\,H^2\right)\,\bar\Pi^2}{\lambda^2}-\frac{1}{2}\,\frac{\left(m^2+H^2\right)\,\bar\Pi^4}{\lambda^4}\label{m11sm}\,\,\,.
\end{eqnarray}

As discussed in section 4, if $\nu^2>0$ then we can bring the $\vec{P}^2$ terms to canonical form by redefining  $P_0\rightarrow P_0\,\gamma-\dot\gamma\,Q_0$, $Q_0\rightarrow Q_0/\gamma$ with $\gamma\equiv 2\,H\nu\sqrt{3}/\lambda^2$. 
The Hamiltonian then keeps the same form as above, however the coefficient of the $P_0^2$ term is now just $1/2$ ({\it i.e.}, $K_{00}=1$), $M_{00}$
\begin{eqnarray}\label{m00new}
M_{00}&=&-\left(\frac{1}{3}+\frac{2\,\lambda^2}{3\,\nu^2}\right)\,\Delta+\frac{\bar\Pi^2}{2\,\nu^2}\,\bar{V}''-\frac{\lambda^2}{\nu^4}\,\bar{V}'{}^2-\frac{H\,\bar\Pi}{2\,\lambda^2\nu^4}\,\left(12\,\lambda^2\,\nu^2+\bar\Pi^4\right)\,\bar{V}'+\\
&&+\frac{1}{32\,\lambda^4\,\nu^4}\,\left[8\left(4m^2-9H^2\right)\lambda^8+12\,\left(m^2-40H^2\right)\,\lambda^6\,\bar\Pi^2+\right.\nonumber\\
&&-\left. 6\,\left(m^2+47H^2\right)\,\lambda^4\bar{\Pi}^4-\left(17m^4+20m^2\,H^2-108 H^4\right)\,\bar\Pi^6-4\,H^2\,\bar\Pi^8\right]\,,\nonumber
\end{eqnarray}
whereas $M_{01}$ goes to $M_{01}/\gamma$ and $V_{10}$ goes to $V_{10}/\gamma$.

%%%%%%%%%%%%%%%%%%%%%%%%%
\section{Removing the $Q_0\,P_1$ term}%%
%%%%%%%%%%%%%%%%%%%%%%%%%

In this appendix we show how to formally bring the Hamiltonian derived in appendix A to the canonical form $\frac{1}{2}p^T\cdot p+\frac{1}{2}q^T\cdot M_{\mathrm {fin}}\cdot q$. We start from the expression 
\begin{eqnarray}
{\cal {H}}=\frac{1}{2}p^T\cdot p+\frac{1}{2}q^T\cdot M\cdot q+p^T\cdot V\cdot q\,\,,\qquad
V=\left(
\begin{array}{cc}
0 & 0 \\
v(t) & 0 
\end{array}\right)
\end{eqnarray}
where $v(t)$ is an arbitrary function of time and $M$ is an arbitrary time-dependent $2\times 2$ ``mass" matrix. We look for a transformation  of the form
\begin{equation}
p\rightarrow A\cdot p+B\cdot q\,\,,\qquad q\rightarrow A\cdot q\,. 
\end{equation}
that can be shown to be canonical if the matrices $A$ and $B$ are such that $A^T\cdot B$ is symmetric. In order to leave the kinetic term untouched, the matrix $A$ has to be orthogonal. Let us write it as
\begin{eqnarray}
A=\left(\begin{array}{cc}
\cos\alpha(t) & \sin\alpha(t) \\
-\sin\alpha(t) & \cos\alpha(t)
\end{array}\right)\,\,.
\end{eqnarray}
The matrix $B$ has to be equal to $\dot{A}-V\cdot A$ to eliminate the cross term $p^T\cdot V\cdot q$ from the Hamiltonian. The function $\alpha(t)$, finally, is determined by the requirement that $A^T\cdot B$ is symmetric. It is easy to prove that such condition is verified if the function $\alpha(t)$ satisfies the differential equation $\dot\alpha(t)=-v(t)/2$. Both $A$ and $B$ are then determined up to an integration constant. By using the equation that relates $A$ to $B$ we find that the Hamiltonian can be brought to the final expression: $\frac{1}{2}p^T\cdot p+\frac{1}{2}q^T\cdot M_{\mathrm {fin}}\cdot q$ with $M_{\mathrm {fin}}=A^T\,M\,A-A^T\,V^T\,V\,A-\ddot{A}^T\,A-A^T\,\dot{V}\,A$.

\bibliographystyle{utphys}
\bibliography{GSHiguchi09}

\end{document}